# Infinite Servers Queue Systems Busy Period Time Length Distribution and Parameters Study through Computational Simulation

**Manuel Alberto M. Ferreira**

Instituto Universitário de Lisboa (ISCTE-IUL), BRU-IUL, ISTAR-IUL, Lisboa, Portugal

**Abstract**

*A FORTRAN program to simulate the operation of infinite servers' queues is presented in this work. Poisson arrivals processes are considered but not only. For many parameters of interest in queuing systems study or application, either there are not theoretical results or, existing, they are mathematically intractable what makes their utility doubtful. In this case a possible issue is to use simulation methods in order to get more useful results. Indeed, using simulation, some experiences may be performed and the respective results used to conjecture about certain queue systems interesting quantities. In this paper this procedure is followed to learn something more about quantities of interest for those infinite servers queue systems, in particular about busy period parameters and probability distributions.*

***Keywords.*** *Infinite servers, queue, Poisson arrivals, simulation, busy period.*

*MSC 2010:* 60G99.

## Introduction

The lack of theoretical results, often difficult to obtain, or its extreme complexity that makes its utility doubtful, leads to the search for numerical methods, in particular simulation methods, in the study of queue systems.

In this work this is done for queue systems with infinite servers. The most famous and studied of these systems is the $M / G / \infty$ queue. For it a lot of results are known and result very clear and simple to use, see for instance (Carrillo, 1981), (Ferreira, 2002) and (Ferreira and Andrade, 2009, 2010, 2010a, 2010b, 2011, 2012).

One of the situations simulated here is related to the consideration of non-Poisson arrivals that is: with inter-arrival times not exponential for which there are no analytical results.

Other results collected are about the busy period[1], for Poisson and non-Poisson arrivals. Indeed, any queue system has a sequence of busy periods and idle periods. A busy period followed by an idle period is a busy cycle. In the study of infinite servers systems busy periods very useful information is, for instance, about the maximum number of customers served simultaneously in a busy period. If something is known about this quantity the system may be dimensioned as a finite servers queue. As there are not analytical results the simulation is one of the issues to study it.

[1] A busy period begins with the arrival of a customer to the system, being it empty, ends when a customer abandons the system, letting it empty, there being always at least one customer in the system.

In the next section, details about the FORTRAN program designed to perform the simulations and the experiences performed are presented. The following section consists in the presentation of the results and the respective comments. The paper ends with a conclusions section.

## The Simulations Proceeding

The simulations were executed using a FORTRAN program, see Appendix, composed of:

*i)* A main program, in FORTRAN language, called FILAESP,

*ii)* A subroutine GERASER,

*iii)* A package SSPLIB,

*iv)* A system function RAN.

The proceedings are as follows:

*i)* The sequential random generation of 25 000 arrivals instants, being the inter-arrivals mean time $\lambda^{-1} = 0.99600$ (a number very close to 1but not exactly 1 because of computational reasons),

*ii)* The generation of 25 000 service times that is added to the arrivals instants so obtaining the departure instants,

*iii)* The ordering of the arrivals and departure instants, through an ordering algorithm making to correspond to each arrival 1 and to each departure -1,

*iv)* The generation, in fact, of the queue adding by order those values 1 and -1, in correspondence with the instants at which they occur,

*v)* The processing of the information, in *iv)*, in order to obtain

*a)* Data related to the state of the system[2]:

- Number of the visits to the assumed states,
- Mean sojourn time in each one of those states.

*b)* Data related to the busy period:

- The maximum number of customers served simultaneously in a busy period,
- The total number of customers served in the busy period,
- The length of the busy period.

The arrivals instant generation is performed in the program FILAESP. The service times in the GERASER subroutine. The arrivals and departures instants ordering are performed in the program FILAESP through the SSPLIB package. The queue building and the processing of the information occur also in FILAESP.

[2] The state of the system, in a given instant, is the number of customers that are being served in that instant.

In the generation of the arrivals and the departures are used sequences of pseudo-random numbers supplied by the system function RAN. In general it is made RAN (E*J), being E constant in each experience and assuming J the values from 1 to 25 000. E was chosen to be an integer with four digits.

- To the arrivals process one or two sequences of pseudo-random numbers are needed as considering M, exponential inter-arrival times, or $E_2$, Erlang with parameter 2 inter-arrival times. In the first case it must be made an option for an integer with four digits, E, and in the second for two integers with four digits that will be designated by E and by F. The same happens with the service distribution, considering so G or G and H, as working with M or $E_2$.

The experiences performed are described below, being $\mu^{-1}$ the mean service time and $\rho = \lambda\mu^{-1}$ the traffic intensity.

- ***M / M / ∞***
  E = 7 528
  F = 7 548
  $\mu^{-1} = 4$
  $\rho = 4.016$
  *Number of observed busy periods*: 208

- ***M / M / ∞***
  E = 7 529
  F = 7 549
  $\mu^{-1} = 5$
  $\rho = 5.020$
  *Number of observed busy periods*: 28

- ***M / $E_2$ / ∞***
  E = 7 528
  G = 7 552
  H = 6 666
  $\mu^{-1} = 4$
  $\rho = 4.016$
  *Number of observed busy periods*: 337

- ***M / $E_2$ / ∞***
  E = 7 529
  G = 6 552
  H = 6 667
  $\mu^{-1} = 5$
  $\rho = 5.020$
  *Number of observed busy periods*: 69

- ***$E_2$* / *$E_2$* / ∞**
  E = 4 536
  F = 4 537
  G = 5 224
  H = 6 225
  $\mu^{-1} = 4$
  $\rho = 4.016$
  *Number of observed busy periods*: 804

- ***$E_2$* / *$E_2$* / ∞**
  E = 4 538
  F = 4 539
  G = 5 228
  H = 6 229
  $\mu^{-1} = 5$
  $\rho = 5.020$
  *Number of observed busy periods*: 208

The mean service times considered, 4 and 5, were those for which a reasonable number of busy periods was obtained, among the highest. In fact, increasing the mean service time the observed busy periods decrease very quickly.

Note that for the systems *M / M / ∞* and *M / $E_2$ / ∞*, for the same values of $\rho$ the arrivals instants generated are identical.

## Results Presentation and Remarks

In Figure 1 and Figure 2 the graphics that represent the mean sojourn times in the various states[3], for the *M / M / ∞* system, considering ρ = 4.016 and ρ = 5.020, respectively, are presented. Beyond the observed mean values the theoretical values are also presented, see (Ramalhoto, 1983), given by

$$T_{M_i} = \frac{\mu^{-1}}{i+\rho}, \qquad i = 0, 1, 2, \ldots \qquad (1).$$

In correspondence with the various states are also indicated the number of times that they were visited, in the right columns.

[3] The state of the system in a given instant is the number of costumers that are being served in that instant.

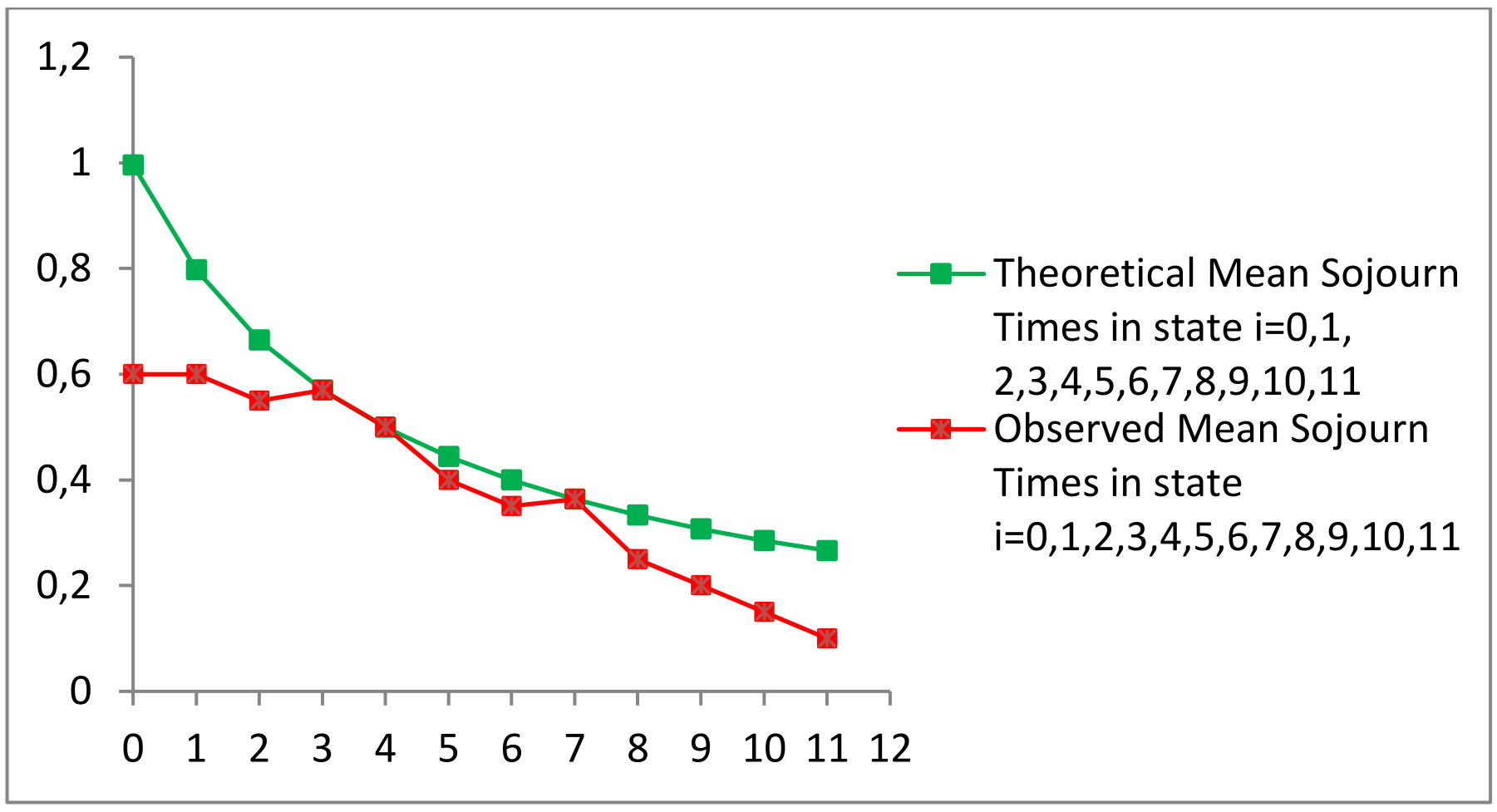

| state | visits |
|---|---|
| 0 | 247 |
| 1 | 2018 |
| 2 | 5970 |
| 3 | 9624 |
| 4 | 10339 |
| 5 | 8809 |
| 6 | 6106 |
| 7 | 3581 |
| 8 | 2132 |
| 9 | 961 |
| 10 | 205 |
| 11 | 7 |

Figure 1: **Mean Sojourn Times, in seconds, theoretical and observed, for the M / M / ∞ queue system in states *i* = 0, 1, …, 11, with ρ = 4.016.**

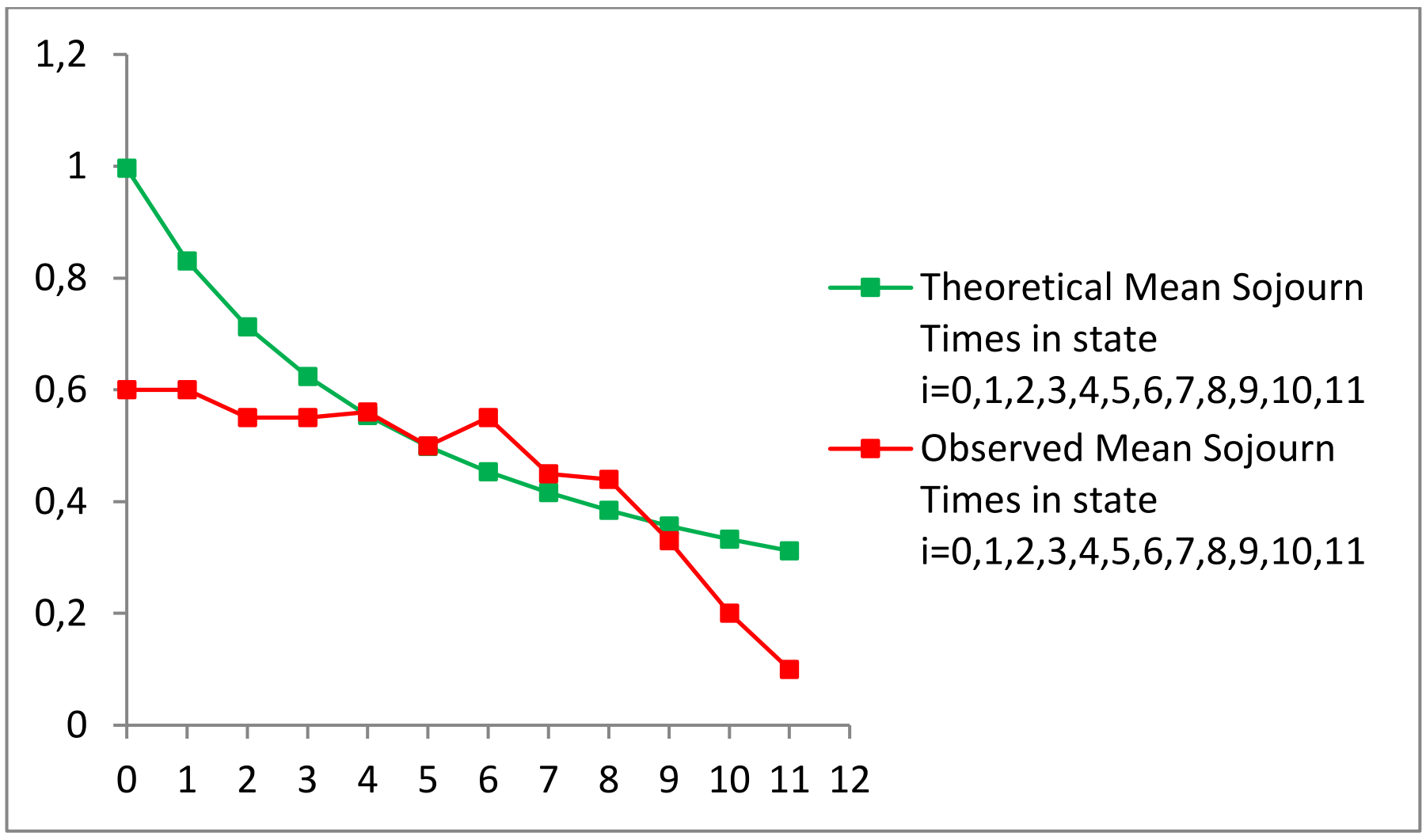

| state | visits |
|---|---|
| 0 | 27 |
| 1 | 441 |
| 2 | 2190 |
| 3 | 5700 |
| 4 | 9267 |
| 5 | 10517 |
| 6 | 9042 |
| 7 | 6207 |
| 8 | 3742 |
| 9 | 2043 |
| 10 | 731 |
| 11 | 92 |

Figure 2: **Mean Sojourn Times, in seconds, theoretical and observed, for the M / M / ∞ queue system in states *i* = 0, 1, …, 11, with ρ = 5.020.**

In Figure 3 and Figure 4 are shown the distributions obtained for the number of customers in the systems M / M / ∞, M / $E_2$ / ∞ and $E_2$ / $E_2$ / ∞ with ρ = 4.016 and ρ = 5.020, respectively. Together is also presented the theoretical distribution, in equilibrium, for the systems M / M / ∞ and M / $E_2$ / ∞, see (Tackács, 1962):

$$p_n = e^{-\rho}\frac{\rho^n}{n!}, \quad n = 0, 1, 2, \dots \qquad (2),$$

obtained performing the direct computations for ρ = 4.016 and ρ = 5.020. E [N] is the mean number of customers in the system.

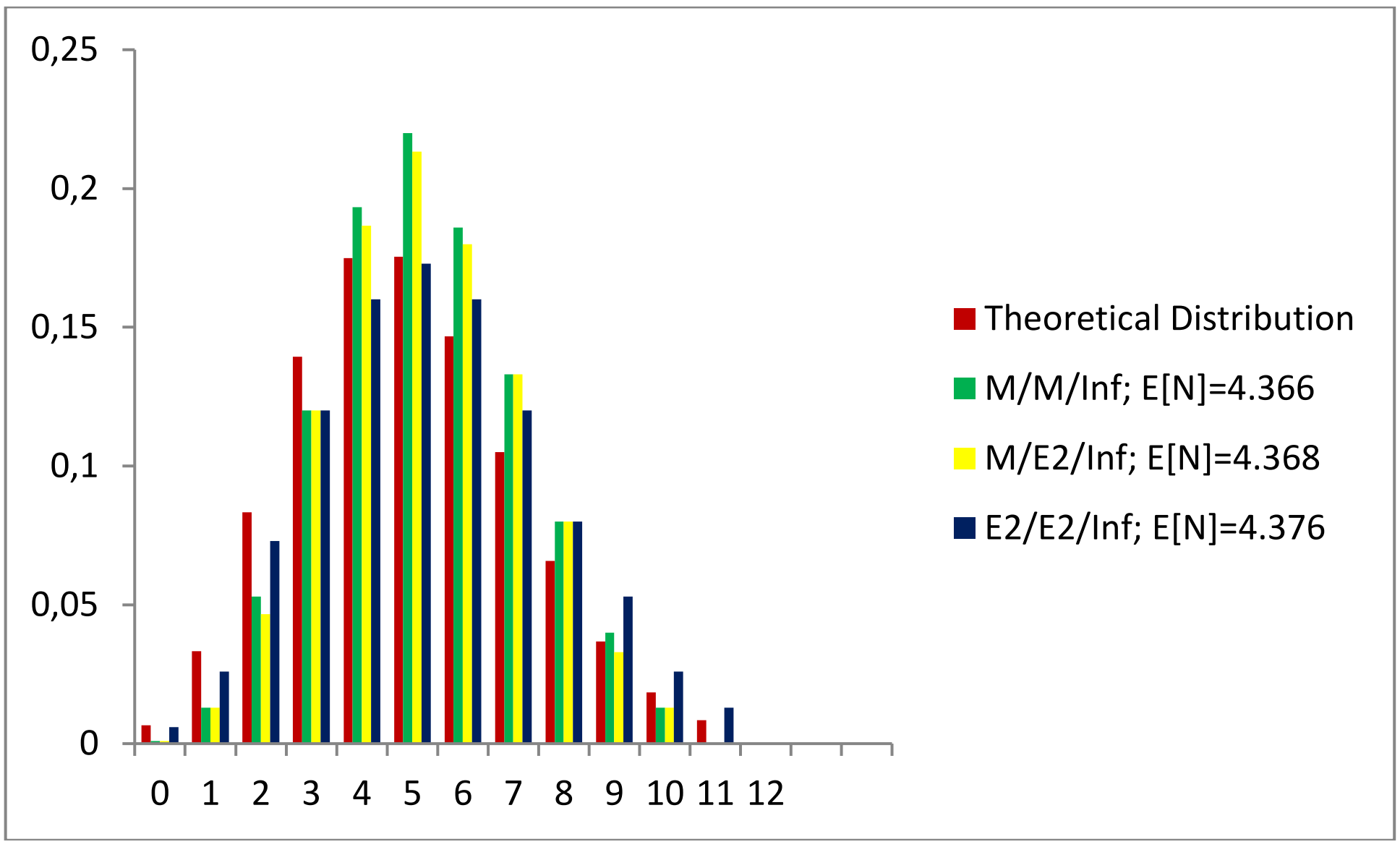

Figure 3: **Distribution of the Number of Customers in the System and Theoretical Distribution for the Systems M / M / ∞, M / $E_2$ / ∞ and $E_2$ / $E_2$ / ∞ with ρ = 4.016.**

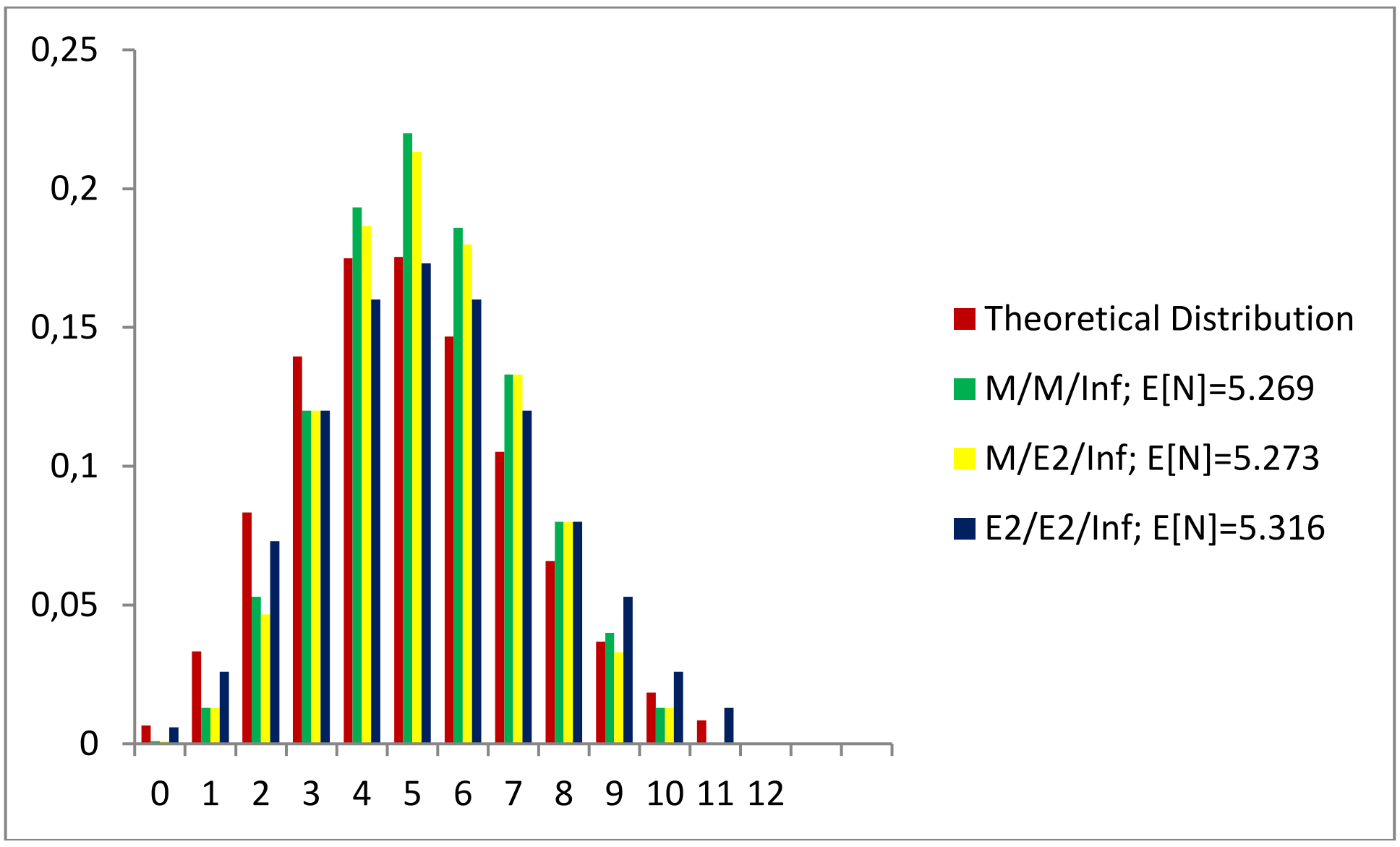

Figure 4: **Distribution of the Number of Customers in the System and Theoretical Distribution for the Systems M / M / ∞, M / $E_2$ / ∞ and $E_2$ / $E_2$ / ∞ with ρ = 5.020.**

These Figures suggest some similarity of behaviour among the empirical distributions and the theoretical distribution. Particularly, in the whole of them the mode is identical to the one of the theoretical distribution. But although for the systems M / M / ∞ and M / $E_2$ / ∞ the empirical distributions are more concentrated around the mode, in comparison with the theoretical distribution, the opposite happens to the system $E_2$ / $E_2$ / ∞. And, surprisingly because for this system it is not known the theoretical distribution, the empirical distributions obtained for $E_2$ / $E_2$ / ∞ seems closer to the theoretical distribution than the ones of the other systems.

As for the differences observed between the systems M / M / ∞ and M / $E_2$ / ∞, for the number of the customers in the system, the adequate interpretation may be as follows: although the systems reach certainly the equilibrium, since the number of the simulated arrivals is quite large there is a strong presence of an initial transitory trend that must last a long time. Note, observing Figure 1 and Figure 2, that the mean sojourn times observed and theoretical, given by (1), for the system M / M / ∞ are quite close. This closeness is better for the states to which corresponds greater frequency.

In Figure 5 and Figure 6, about the maximum number of customers served simultaneously in the busy period, it is remarked great diversity in the distributions form. It is always observed a great frequency for the state 1. In the $E_2$ / $E_2$ / ∞ infinite systems it is always the mode. Curiously, these systems being able to serve any number of customers, present, in these simulations, few customers being served simultaneously: never above the number 14, only assumed by $E_2$ / $E_2$ / ∞ infinite systems. This fact is acceptable, in terms of the theoretical distribution, since in the Poisson distribution the values greater than the mode, far away from it, are little probable. Note still that, excluding from this analysis the state 1, the distributions of maximum number of customers served simultaneously in the $E_2$ / $E_2$ / ∞ infinite systems busy period are more scattered than those of the others. This is in accordance with the fact that, for the same number of arrivals, much more busy periods are observed for the $E_2$ / $E_2$ / ∞ infinite systems.

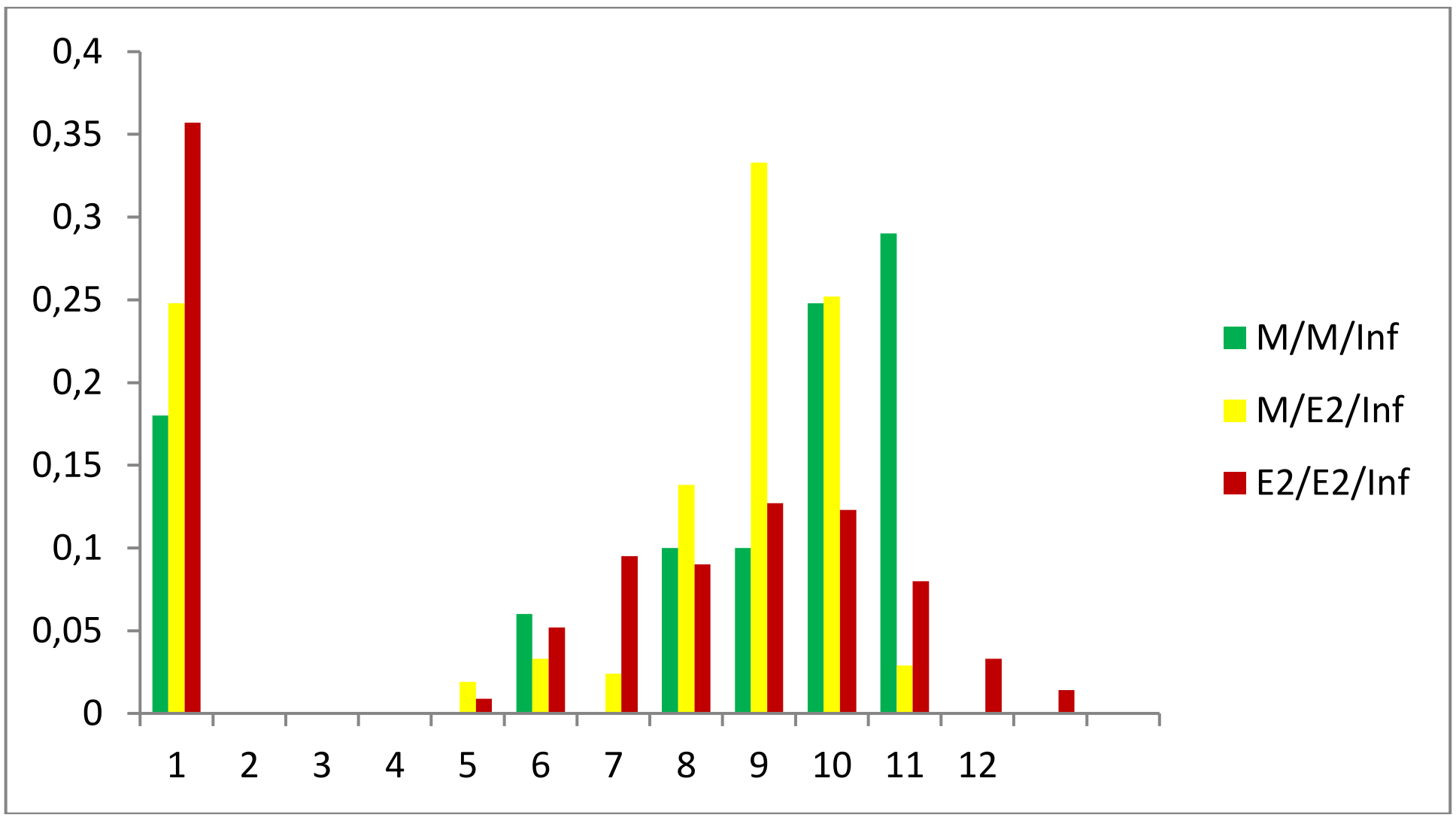

Figure 5: **Distribution of the Maximum Number of Customers Served Simultaneously in a Busy Period for the Systems M / M / ∞, M / $E_2$ / ∞ and $E_2$ / $E_2$ / ∞ with ρ = 4.016.**

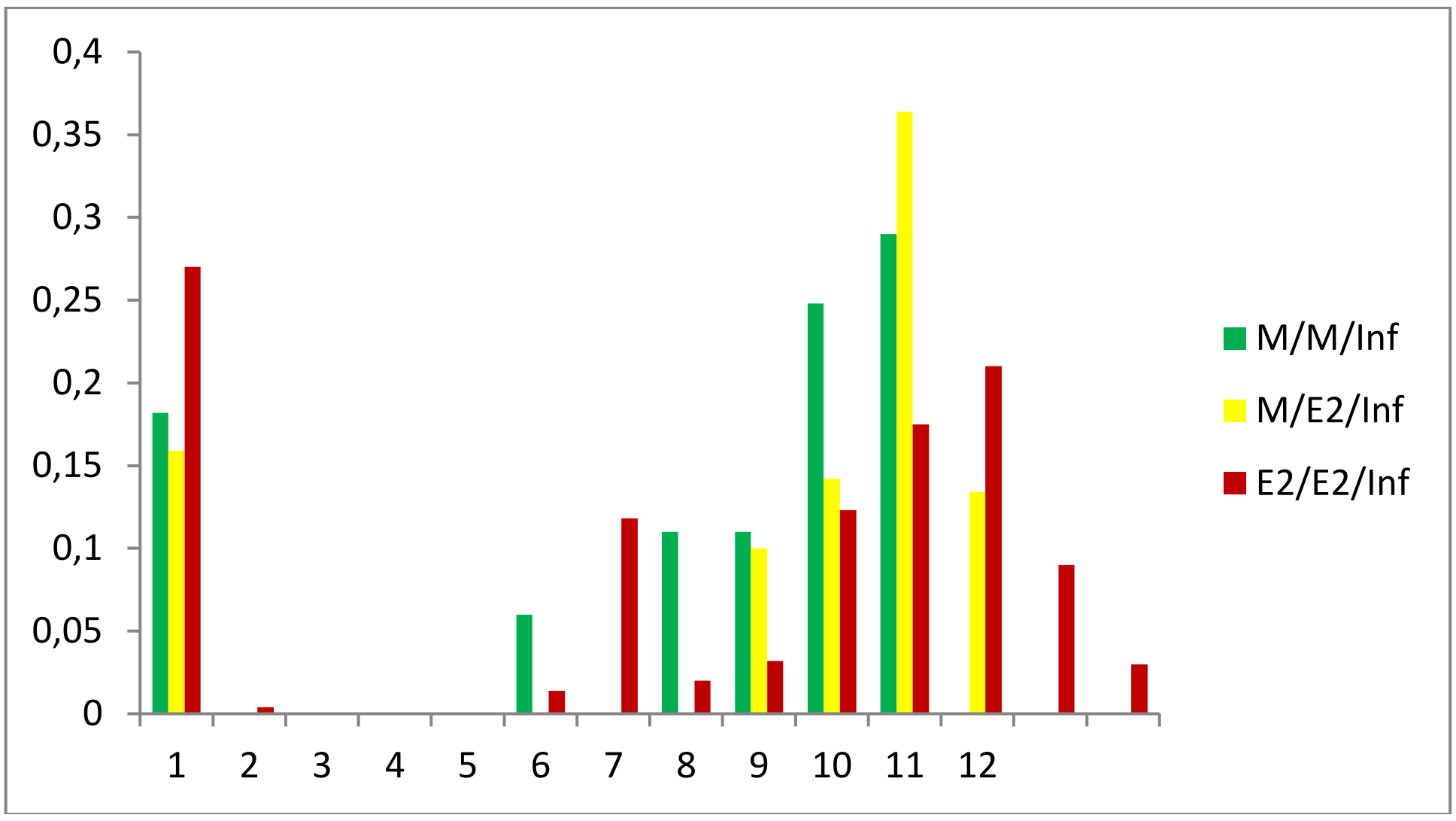

Figure 6: **Distribution of the Maximum Number of Customers Served Simultaneously in a Busy Period for the Systems M / M / ∞, M / $E_2$ / ∞ and $E_2$ / $E_2$ / ∞ with ρ = 5.020.**

Figure 7 suggests a more and more smooth behaviour of the busy period lengths distribution frequency curve when going from M / M / ∞ system for the M / $E_2$ / ∞ system and then for the $E_2$ / $E_2$ / ∞ system. The whole of them present a great frequency concentration for the lowest values of the busy period lengths but, in the case of M / M / ∞ system, the interval along which the observations spread has more than the double of the length of the other systems. Remember, again, that for the M / M / ∞ system much

**ρ = 4.016**

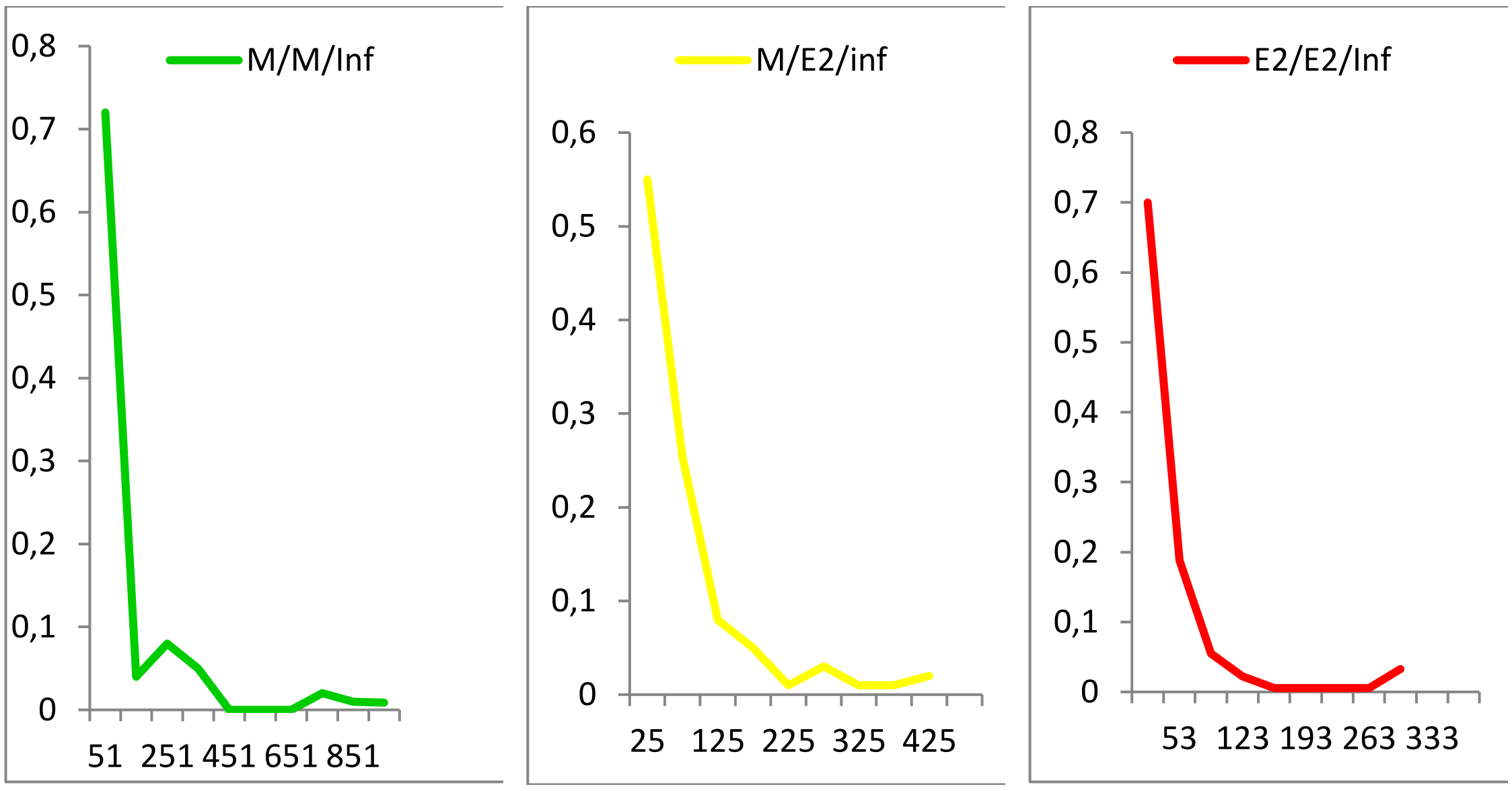

Figure 7: **Distribution of Observed Busy Period Lengths, in seconds, for the Systems M / M / ∞, M / $E_2$ / ∞ and $E_2$ / $E_2$ / ∞ with ρ = 4.016.**

less busy periods are observed than for the M / $E_2$ / ∞ system and, for this one less than for the $E_2$ / $E_2$ / ∞ system. Then the curve of frequencies for the M / M / ∞ system spreads along that interval with two deep valleys. The one of the M / $E_2$ / ∞ system presents also two valleys, but less deep, and in $E_2$ / $E_2$ / ∞ system practically they are not observed. Note, also that those valleys occur for different values in the M / M / ∞ and M / $E_2$ / ∞ systems.

Figure 8 suggests a greater similarity among the busy period lengths distributions of the various systems. Again it is observed a great concentration in the lowest values,

**ρ = 5.020**

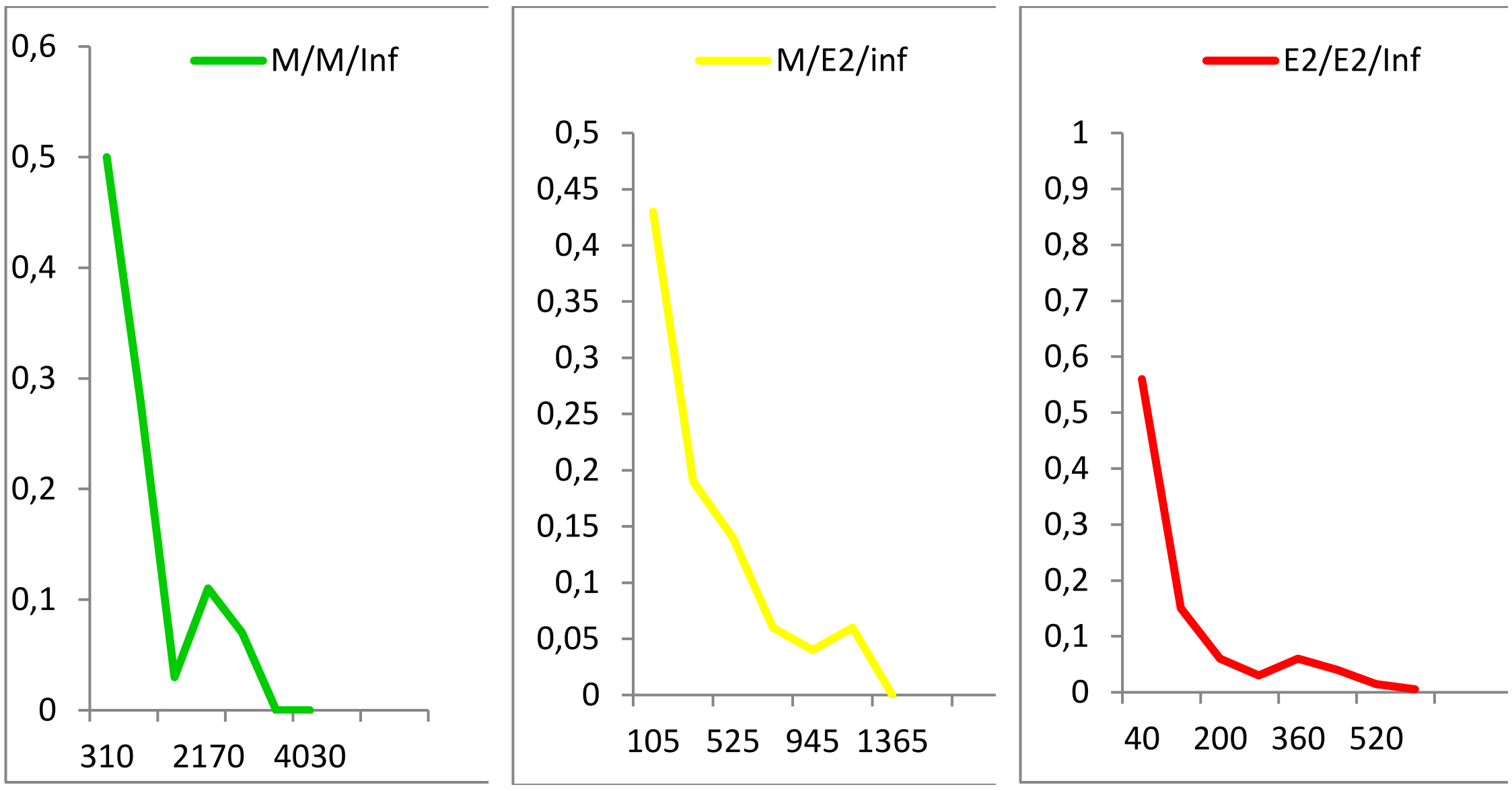

Figure 8: **Distribution of Observed Busy Period Lengths, in seconds, for the Systems M / M / ∞, M / $E_2$ / ∞ and $E_2$ / $E_2$ / ∞ with ρ = 5.020.**

although lesser than in Figure 7. But now the whole of them present one valley, ocurring for different values. Still goes on observing a great disparity among the maximum values assumed by the busy period lengths.

It seems obvious, either in Figure 7 or in Figure 8, a sharp observations lack in the intermediate values zone for the busy period lengths.

Note also that the Figures 7 and 8 are in accordance with the studies that point in order that the busy period length distribution of the M / G / ∞ system is right asymmetric and leptokurtic see (Ferreira and Ramalhoto, 1994).

Call $X$ and $Y$ the maximum number of customers served simultaneously and the total number of served customers, respectively, in the busy period. Performing the regression of $Z = lnY$ on $X$, the results obtained are presented graphically in Figure 9 and the most interesting fact is, maybe, the similarity of the lines behaviour, for the various systems, for the two values of ρ considered. In fact, among the whole considered systems there is a great resemblance of the values of $\hat{\alpha}$ (the intercept) and $\hat{\beta}$(the slope) for the two values of ρ considered. Actually, it seems natural that the relation between $Z$ and $X$ does not depend on ρ. The values of ρ will only influence the values of $Z$ and $X$ that may occur and not the relation between them. Otherwise, will it be true that the differences observed in the values of $\hat{\alpha}$ and $\widehat{\beta}$, for the various systems, not being too great, allows facing the hypothesis that the relation between $Z$ and $X$ is identical for those systems? Maybe yes if it is paid attention to the similarities observed in its behaviour, namely the ones related with the distribution of the number of customers in the various systems.

| ρ = 4.016 | ρ = 5.020 |
|---|---|

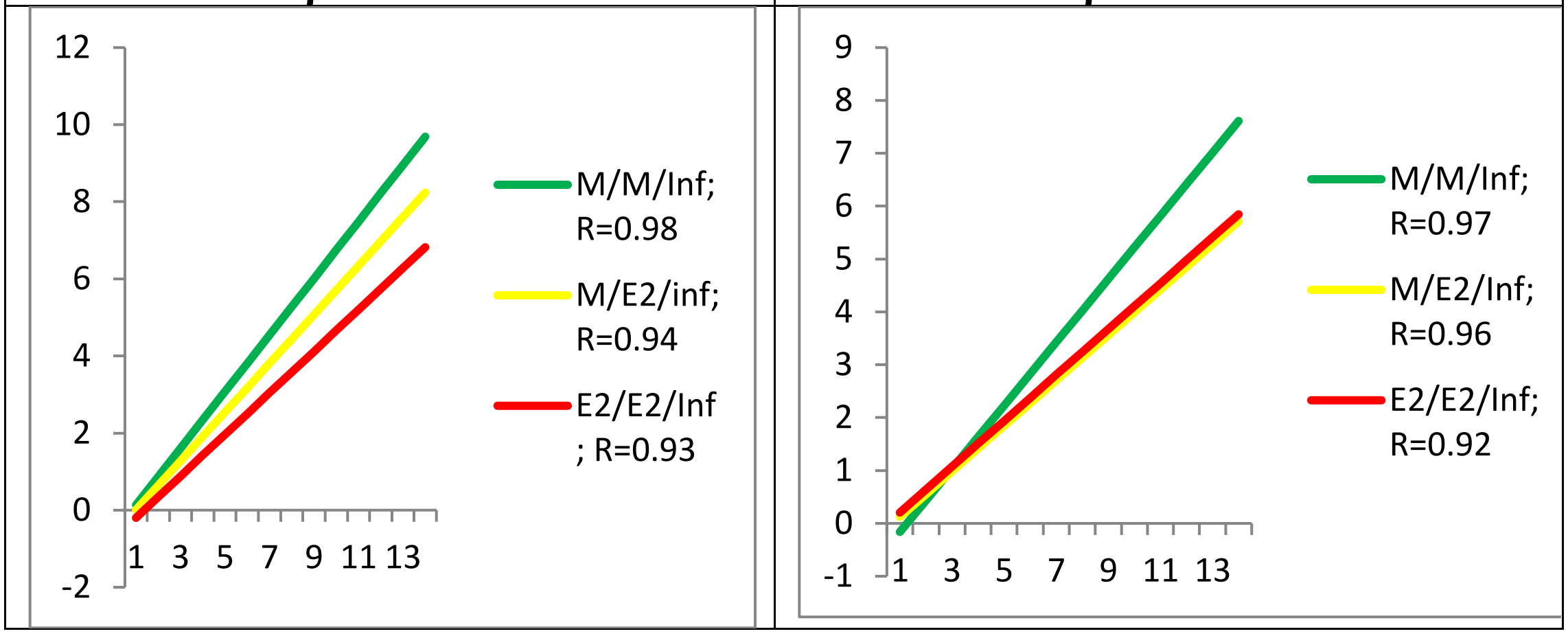

Figure 9: **Regression of $Z$ over $X$, for the Systems M / M / ∞, M / $E_2$ / ∞ and $E_2$ / $E_2$ / ∞ with ρ = 4.016 and ρ = 5.020. R is the Linear Correlation Coefficient.**

## Concluding Remarks

It is manifest the great waste of these systems when looking to the maximum number of customers present simultaneously in the system.

It is uncontroversial, also, that there is a strong exponential relation between the maximum number of customers served simultaneously in a busy period and the total number of served customers. The question is if either it is always the same or how will it change either with the values of ρ or from system to system.

On the busy period it is important to note also:

*i*) The great occurrence of busy periods with only one served customer,

*ii*) The great amplitude of the interval at which occur the values of the lengths of the busy periods, although with a great irregularity, and a great occurrence of low values.

The results of these simulations seem to suggest also that the systems GI /G / ∞ may be quite well approximated by systems M / G / ∞, at least when the GI process possesses a regularity not very far from the one of the Poisson process.

## Appendix

```
            PROGRAM FILAESP

C     APAGUE 666 OU 555 CONFORME QUEIRA TEMPO INTERCHEGADAS
C      EXPONENCIAL OU ERLANG DE PARAMETRO 2
      DIMENSION V(25000),Y1(25000),Y2(25000),TEPSER(25000)
      DIMENSION C(25000),AUX(50000),P(25000),N(50000),F(50000)
      DIMENSION Z(50000),VAL(50000),ARG(50000)
      DIMENSION NO(0:50000),T(0:25000),TM(0:25000)
      DIMENSION TTR(0:25000),TMR(0:25000)
      DIMENSION BUPE(0:25000),NA(0:25000)
      WRITE(*,*)'O CODIGO DOS SERVICOS E: 0 PARA A PARETO, 1 PARA A'
      WRITE(*,*)'EXPONENCIAL, 2 PARA A ERLANG, 3 PARA A LOGNORMAL,'
      WRITE(*,*)'4 PARA A MISTURA DE EXPONENCIAIS COM RPARAMETRO,'
      WRITE(*,*)'5 PARA A MISTURA DE ERLANG.'
      WRITE(*,*)' '
      WRITE(*,*)' QUAL E O CODIGO DA DISTRIBUICAO DE SERVICO?'
      READ(*,*) ICOD
      WRITE(*,*)' '
      WRITE(*,*)' BOA SORTE NA VIAGEM AO MUNDO DA SIMULACAO '
      WRITE(*,*)' '
      WRITE(*,*)' '
      U=0.99600
      DO 100 I=1,25000
555         V(I)=ALOG(RAN(E*I))*(-U/2.0)
666         V(I)=ALOG(RAN(E*I))*(-U/2.0)+ALOG(RAN(F*I))*(-U/2.0)
100    CONTINUE
      CALL GERASER(TEPSER)
      C(1)=V(1)
      Z(1)=C(1)
      F(1)=1
      Z(25001)=V(1)+TEPSER(1)
      F(25001)=-1
      DO 500 I=2,25000
            C(I)=C(I-1)+V(I)
            P(I)=C(I)+TEPSER(I)
            Z(I)=C(I)
            F(I)=1
            Z(I+25000)=P(I)
            F(I+25000)=-1
500    CONTINUE
      X=0.0
      ICOL=1
      IROW=50000
      NDIM=50000
      CALL ATSG(X,Z,F,AUX,IROW,ICOL,ARG,VAL,NDIM)
      N(1)=1
      DO 600 I=2,50000
            N(I)=N(I-1)+VAL(I)
600   CONTINUE
      MAX=1
      DO 650 I=2,50000
            IF (N(I).GE.MAX)MAX=N(I)
650   CONTINUE
      WRITE(10,*)' SIMULACAO FILA DE ESPERA M|G|□□‘
      DO 700 K=0,MAX
      WRITE(10,*)' TEMPOS DE RECORRENCIA DO ESTADO ‘,K
      J=1
      NO(K)=0
      T(K)=0.0
      TM(K)=0.0
      DO 660 I=1,49999
```

```
                IF(N(I).EQ.K) THEN
                NO(K)=NO(K)+1
                T(K)=T(K)+(ARG(I+1)-ARG(I))
                J=J+1
                ENDIF
660     CONTINUE
        DO 1000 J=1,NO(K)-1
                TTR(K)=TTR(K)
1000    CONTINUE
        IF(NO(K).NE.0)TM(K)=T(K)/NO(K)
        IF(NO(K).GT.1)TMR(K)=TTR(K)/(NO(K)-1)
        IF(NO(K).EQ.1)TMR(K)=0
        WRITE(10,*)'ESTADO',K
        WRITE(10,*)'NUMERO DE VISITAS =',NO(K)
        WRITE(10,*)'TEMPO DE PERMANENCIA =',T(K)
        WRITE(10,*)'TEMPO MEDIO DE PERMANENCIA =',TM(K)
        WRITE(10,*)'TEMPO TOTAL DE RECORRENCIA =',TTR(K)
        WRITE(10,*)'TEMPO MEDIO DE RECORRENCIA =',TMR(K)
700     CONTINUE
        TTBUPE=ARG(50000)-ARG(1)-T(0)
        NTBUPE=1+N0(0)
        TMBUPE=TTBUPE/NTBUPE
        TTIDP=T(0)
        NTIDP=NO(0)
        TMIDP=TM(0)
        WRITE(10,*)'NO TOTAL DE PERIODOS DE OCUPACAO =',NTBUPE
        WRITE(10,*)'TEMPO TOTAL DE BUSY PERIOD =',TTBUPE
        WRITE(10,*)'TEMPO MEDIO DE BUSY PERIOD =',TMBUPE
        WRITE(10,*)'NO TOTAL DE PERIODOS DE DESOCUPACAO =',NTIDP
        WRITE(10,*)'TEMPO TOTAL DE IDLE PERIOD =',TTIDP
        WRITE(10,*)'TEMPO MEDIO DE IDLE PERIOD =',TMIDP
        BUPE(0)=0
        NA(0)=1
        NP=0
        DO 5000 I=1,49999
                IF(N(I).EQ.0) THEN
                NP=NP+1
                WRITE(10,*)'BUSY PERIOD NUMERO',NP
                NA(NP)=I
                NB=0
        DO 4000 J=NA(NP-1),NA(NP)-1
                IF(N(J+1).GT.N(J))NB=NB+1
4000     CONTINUE
        IF(NP.EQ.1)NB=NB+1
        WRITE(10,*)'NUMERO DE CLIENTES ATENDIDOS=',NB
        MAXI=1
        DO 3900 J=NA(NP-1),NA(NP)
                IF(N(J).GE.MAXI) MAXI=N(J)
3900    CONTINUE
        WRITE(10,*)'NUMERO MAXIMO DE CLIENTES ATENDIDOS SIMULTANEAMENTE=',MAXI
        BUPE(NP)=ARG(I+1)
        RBUPE=BUPE(NP)-BUPE(NP-1)-ARG(I+1)+ARG(I)
        IF(NP.EQ.1)RBUPE=BUPE(1)-ARG(I+1)+ARG(I)-ARG(1)
        WRITE(10,*)'COMPRIMENTO=',RBUPE
        ENDIF
5000    CONTINUE
        NP=NP+1
        WRITE(10,*)'BUSY PERIOD NUMERO=',NP
        NB=0
        DO 4001 J=NA(NP-1),49999
                IF(N(J+1).GT.N(J))NB=NB+1
4001     CONTINUE
        IF(NP.EQ.1)NB=NB+1
        WRITE(10,*)'NUMERO DE CLIENTES ATENDIDOS='NB
        MA=1
        DO 4002 J=NA(NP-1),50000
                IF(N(J).GE.MA)MA=N(J)
4002    CONTINUE
        WRITE(10,*)'NUMERO MAXIMO DE CLIENTES ATENDIDOS SIMULTANEAMENTE='MA
        RBUPE=ARG(50000)-BUPE(NP-1)
        IF(NP.EQ.1)RBUPE=ARG(50000)-ARG(1)
        WRITE(10,*)'COMPRIMENTO= ',RBUPE
        END
```

**SUBROUTINE GERASER (T)**

```
      DIMENSION T(25000)
                  ICOD=2
      IF (ICOD.EQ.0) THEN
      PRINT*, 'INTRODUZA O VALOR DO COEFICINTE DE VARIACAO'
      READ*,GAMA
      ALFA=2*GAMA/(GAMA-1.0)
      RK=E*(GAMA+1.0)/(2*GAMA)
      ELSEIF (ICOD.EQ.4) THEN
      PRINT*,'INTRODUZA O VALOR DO PARAMETRO DA MISTURA'
      READ(*,*)RPARAMETRO
      ENDIF
      IX=35
      IY=43
      IIX=9
      IJX=5
      IKX=11
      ILX=13
      DO 10 I=1,250000
                  CALL RANDU(IX,IY,FL)
                  IF(ICOD.EQ.O) THEN
                  T(I)=RK/(1-FL)**(1.0/ALFA)
                  ELSE IF (ICOD.EQ.1) THEN
                  T(I)=-7.0*ALOG(RAN(G*I))
                  ELSE IF (ICOD.EQ.2) THEN
      T(I)=-(4.0/2.0)*ALOG(RAN(G*I))-(4.0/2.0)*ALOG(RAN(H*I))
                  ELSE IF (ICOD.EQ.3) THEN
                  CALL RANDU(JX,IJY,YFFL)
                  IJX=IJY
                  T(I)=EXP((-2*ALOG(YFFL))**(0.5)*COS(8*ATAN(1.0)*FL))
                  ELSE IF (ICOD.EQ.4) THEN
                  CALL RANDU (IKX,IKY,RFL)
                  IKX=IKY
                  CALL RANDU (ILX,ILY,SFL)
                  ILX=ILY
                  IF (FL.LE.RPARAMETRO) THEN
                  T(I)=-(3.45/2.0)*(1.0/RPARAMETRO)*LOG(RFL)
                  ELSE
                  T(I)=-(3.45/2.0)*(1.0/1.0-RPARAMETRO)*LOG(SFL)
                  ENDIF
                  ELSE IF (ICOD.EQ.5) THEN
                  CALL RANDU (IKX,IKY,RFL)
                  IKX=IKY
                  CALL RANDU (ILX,ILY,SFL)
                  ILX=ILY
                  CALL RANDU (IMX,IMY,TFL)
                  IMX=IMY
                  CALL RANDU (INX,INY,UFL)
                  INX=INY
                  CALL RANDU (IPX,IPY,VFL)
                  IPX=IPY
                  CALL RANDU (IQX,IQY,WFL)
                  IQX=IQY
                  IF(RFL.LT.0.400) THEN
      T(I)=-((10.0/2.7)/4.0)*(LOG(UFL)+LOG(SFL)+LOG(TFL)+LOG(WFL))
                  ELSE IF ((RFL.GT.0.4000).AND.(RFL.LT.0.75)) THEN
                  T(I)=-((10.0/4.2)/2.0)*(LOG(VFL)+LOG(WFL))
                  ELSE IF (RFL.GT.0.75) THEN
                  T(I)=-((10.0/3.6)/3.0)*(LOG(VFL)+LOG(WFL)+LOG(SFL))
                              ENDIF
                  ENDIF
10                CONTINUE
      END
```